\documentclass[showpacs,preprintnumbers,amsmath,amssymb]{revtex4}

\usepackage{graphicx}% Include figure files
\usepackage{dcolumn}% Align table columns on decimal point
\usepackage{bm}% bold math

\begin{document}

%\preprint{APS/123-QED}

%\title{Manuscript Title:\\with Forced Linebreak}% Force line breaks with \\
\title{Diffusion-controlled phase growth on dislocations
\footnote{An extended and  revised version of
the paper presented in MS\&T'08, October 5-9, 2008, Pittsburgh,
Pennsylvania, USA.}} % Force line breaks with \\
\author{ A. R. Massih}%
\affiliation{%
Quantum Technologies, Uppsala Science Park, SE-751 83 Uppsala, Sweden and \\
Malm\"{o} University, SE-205 06 Malm\"{o}, Sweden}

\date{\today}% It is always \today, today,

\begin{abstract}
We treat the problem of diffusion of solute atoms around screw
dislocations. In particular, we express and solve the diffusion
equation, in radial symmetry, in an elastic field of a screw
dislocation subject to the flux conservation boundary condition at the
interface of a new phase. We consider an incoherent second-phase
precipitate growing under the action of the stress field of a screw
dislocation. The second-phase growth rate as a function of the
supersaturation and a strain energy parameter is evaluated in spatial
dimensions $d=2$ and $d=3$. Our calculations show that an increase in
the amplitude of dislocation force, e.g. the magnitude of the Burgers
vector, enhances the second-phase growth in an alloy. Moreover, a
relationship linking the supersaturation to the precipitate size in
the presence of the elastic field of dislocation is
calculated.\bigskip

\end{abstract}

\maketitle

\section{Introduction}
\label{sec:intro}
Dislocations can alter different stages of the precipitation process
in crystalline solids, which consists of nucleation, growth and
coarsening \cite{Larche_1979,Wagner_Kampmann_1991}. Distortion of the
lattice in proximity of a dislocation can enhance nucleation in
several ways \cite{Porter_Easterling_1981,Christian_2002}. The main
effect is the reduction in the volume strain energy associated with
the phase transformation. Nucleation on dislocations can also be
helped by solute segregation which raises the local concentration of
the solute in the vicinity of a dislocation, caused by migration of
solutes toward the dislocation, the Cottrell atmosphere effect. When
the Cottrell atmosphere becomes supersaturated, nucleation of a new
phase may occur followed by growth of nucleus. Moreover, dislocation
can aid the growth of an embryo beyond its critical size by providing
a diffusion passage with a lower activation energy.

Precipitation of second-phase along dislocation lines has been
observed in a number of alloys
\cite{Aaronson_et_al_1971,Aaron_Aaronson_1971}. For example, in
Al-Zn-Mg alloys, dislocations not only induce and enhance nucleation
and growth of the coherent second-phase MgZn$_2$ precipitates, but
also produce a spatial precipitate size gradient around them
\cite{Allen_Vandesande_1978,Deschamps_et_al_1999,Deschamps_Brechet_1999}.
Cahn \cite{Cahn_1957} provided the first quantitative model for
nucleation of second-phase on dislocations in solids.  In Cahn's
model, it is assumed that a cross-section of the nucleus is circular,
which is strictly valid for a screw dislocation
\cite{Larche_1979}. Also, it is posited that the nucleus is incoherent
with the matrix so that a constant interfacial energy can be allotted
to the boundary between the new phase and the matrix. An incoherent
particle interface with the matrix has a different
atomic configuration than that of the phases. The matrix is
an isotropic elastic material and the formation of the precipitate
releases the elastic energy initially stored in its volume. Moreover,
the matrix energy is assumed to remain constant by precipitation. In
this model, besides the usual volume and surface energy terms in the
expression for the total free energy of formation of a nucleus of a
given size, there is a term representing the strain energy of the
dislocation in the region currently occupied by the new phase. Cahn's
model predicts that both a larger Burgers vector and
a more negative chemical free energy change between the precipitate
and the matrix induce higher nucleation rates, in agreement with
experiment \cite{Aaronson_et_al_1971,Aaron_Aaronson_1971}.

Segregation phenomenon around dislocations, i.e. the Cottrell
atmosphere effect, has been observed among others in Fe-Al alloys
doped with boron atoms \cite{Blavette_et_al_1999} and in silicon
containing arsenic impurities \cite{Thompson_et_al_2007}, in
qualitative agreement with Cottrell and Bilby's predictions
\cite{Cottrell_Bilby_1949}. Cottrell and Bilby considered segregation
of impurities to straight-edge dislocations with the Coulomb-like
interaction potential of the form $\phi=A\sin \theta/r$, where $A$
contains the elasticity constants and the Burgers vector, and
$(r,\theta)$ are the polar coordinates. Cottrell and Bilby ignored the
flow due concentration gradients and solved the simplified diffusion
equation in the presence of the aforementioned potential field. The
model predicts that the total number of impurity atoms removed from
solution to the dislocation increases with time $t$ according to $N(t)
\sim t^{2/3}$, which is good agreement with the early stages of
segregation of impurities to dislocations, e.g. in iron containing
carbon and nitrogen \cite{harper_1951}. A critical review of the
Bilby-Cottrell model, its shortcomings and its improvements are given
in \cite{Bullough_Newman_1970}.

The object of our present study is the diffusion-controlled growth of
a new phase, i.e., a post nucleation process in the presence of
dislocation field rather than the segregation effect. As in Cahn's
nucleation model \cite{Cahn_1957}, we consider an incoherent
second-phase precipitate growing under the action of a screw
dislocation field.  This entails that the stress field due to
dislocation is pure shear. The equations used for diffusion-controlled
growth are radially symmetric. These equations for second-phase in a solid or from a
supercooled liquid have been, in the absence of an external field, solved
by Frank \cite{Frank_1950} and discussed by Carslaw and Jaeger
\cite{Carslaw_Jaeger_1959}. The exact analytical solutions of the
equations and their various approximations thereof have been
systematized and evaluated by Aaron et al. \cite{Aaron_et_al_1970},
which included the relations for growth of planar
precipitates. Applications of these solutions to materials can
be found in many publications, e.g. more recent papers on growth of
quasi-crystalline phase in Zr-base metallic glasses
\cite{Koster_et_al_1996} and growth of Laves phase in Zircaloy
\cite{Massih_et_al_2003}.  We should also mention another
theoretical approach to the problem of nucleation and growth of an
incoherent second-phase particle in the presence of dislocation
field \cite{Sundar_Hoyt_1991}. Sundar and Hoyt 
\cite{Sundar_Hoyt_1991} introduced the dislocation field, as in Cahn
\cite{Cahn_1957}, in the nucleation part of the model,
while for the growth part the steady-state solution of the concentration field
(Laplace equation) for elliptical particles was utilized.

The organization of this paper as follows. The formulation of the
problem, the governing equations and the formal solutions are given in
section \ref{sec:formul}. Solutions of specific cases are presented in
section \ref{sec:comp}, where the supersaturation as a function of the
growth coefficient is evaluated as well as the spatial variation of
the concentration field in the presence of dislocation. In section
\ref{sec:disc}, besides a brief discourse on the issue of interaction
between point defects and dislocations, we calculate the
size-dependence of the concentration at the curved precipitate/matrix
for the problem under consideration. We have carried out our
calculations in space dimensions $d=2$ and $d=3$.  Some mathematical
analyses for $d=3$ are relegated to appendix \ref{sec:appa}.

\section{Formulation and general solutions}
\label{sec:formul}

  We consider the problem of growth of the new phase, with radial
symmetry (radius $r$), governed by the diffusion of a single entity,
$u\equiv u(r,t)$, which is a function of space and time $(r,t)$. $u$
can be either matter (solvent or solute) or heat (the latent heat of
formation of new phase). The diffusion in the presence of an external
field obeys the Smoluchowski equation \cite{Chandra_1943} of the form
\begin{eqnarray}
  \label{eqn:smolu}
 \frac{\partial u}{\partial
t} & = & \nabla \cdot \mathbf{J}, \\
\label{eqn:smolu-flux}
 \mathbf{J} & = & D(\nabla u-\beta\mathbf{F}u),
\end{eqnarray}
\noindent
where $D$ is the diffusivity, $\beta=1/k_BT$, $k_B$ the Boltzmann
constant, $T$ the temperature, and $\mathbf{F}$ is an external field
of force. The force can be local (e.g., stresses due to dislocation
cores in crystalline solids) or caused externally by an applied field (e.g., electric field
acting on charged particles). If the acting force is conservative, it
can be obtained from a potential $\phi$ through $\mathbf{F}=-\nabla
\phi$. The considered geometric condition applies
to the case of second-phase particles growing in a
solid solution under phase transformation \cite{Massih_et_al_2003} or
droplets growing either from vapour or from a second liquid
\cite{Frank_1950}. A steady state is reached when
$\mathbf{J}=\mathrm{const.}=0$, resulting in
$u=u_0\exp(-\beta\phi)$.

Here, we suppose that the diffusion field is along
the core of dislocation line and that a cross-section of the
precipitate (nucleus), perpendicular to the dislocation, is circular,
i.e., the precipitate surrounds the dislocation. Furthermore, we treat
the matrix and solution as linear elastic isotropic media. The elastic
potential energy of a stationary dislocation of length $l$ is given by
\cite{Kittel_1996,Friedel_1967}
\begin{equation}
  \label{eqn:screw}
  \phi = A\ln\frac{r}{r_0}, \qquad \qquad \qquad \textrm{for}\quad r\ge r_0
\end{equation}
\noindent
where $A=Gb^2l/4\pi$ for screw dislocation, $G$ is the elastic shear
modulus of the crystal, $b$ the magnitude of the Burgers vector, $\nu$
Poisson's ratio, and $r_0$ is the usual effective core radius. Also,
we assume that the dislocation's elastic energy is relaxed within the
volume occupied by the precipitate and that the precipitate is
incoherent with the matrix. Hence the interaction energy between the
elastic field of the screw dislocation and the elastic field of the solute
is zero. In the case of an edge dislocation and coherent
precipitate/matrix interface, this interaction is non-negligible.

We study the effect of the potential field (\ref{eqn:screw}) on
diffusing atoms in solid solution using the Smoluchowski
equation (\ref{eqn:smolu}). The governing
equation in spherical symmetry, in $d$ spatial dimension, with $B \equiv \beta A$, is 
\begin{equation}
  \label{eqn:pde-1}
  \frac{1}{D}\frac{\partial u}{\partial t}=
  \frac{\partial^2 u}{\partial r^2}+(d-1+B)\frac{1}{r}\frac{\partial
u}{\partial r}+(d-2)B\frac{u}{r^2}.
\end{equation}
\noindent
 Making a usual change of variable to the
dimensionless reduced radius $s=r/\sqrt{Dt}$, the partial differential
equation (\ref{eqn:pde-1}) is reduced to an ordinary differential
equation of the form
\begin{equation}
  \label{eqn:ode-1}
  \frac{{\rm d^2} u}{{\rm d}s^2}+\Big(\frac{s}{2}+\frac{d-1+B}{s}\Big)\frac{{\rm
d}u}{{\rm d}s}
+(d-2)B\frac{u}{s^2}=0,
\end{equation}
\noindent
with the boundary conditions, $u(\infty) = u_m$, and $u(2\lambda) =
u_s$, where $u_m$ is the mean (far-field) solute concentration in the
matrix and $u_s$ is the concentration in the matrix at the
new-phase/matrix interface determined from thermodynamics of new
phase, i.e., phase equilibrium and the capillary effect.  Moreover,
the conservation of flux at the interface radius $R=2\lambda\sqrt{Dt}$ gives
\begin{equation}
  \label{eqn:flux-1}
  K_d R^{d-1}\vert \mathbf{J}\vert_{r=R}  = q \frac{{\rm d}V_d}{{\rm d}t},
\end{equation}
\noindent
where $K_d=2\pi^{d/2}/\Gamma(d/2)$, $\Gamma(x)$ the usual
$\Gamma$-function, $V_d=2\pi^{d/2}R^d/d\Gamma(d/2)$, and  $q$  the
amount of the diffusing entity ejected at the boundary of the growing
phase per unit volume of the latter (new phase) formed. In $s$-space,
equation (\ref{eqn:flux-1}) is written as  
\begin{equation}
  \label{eqn:conserve-flux}
  \Big(\frac{{\rm d}u}{{\rm d}s}\Big)_{s=2\lambda} = -\Big(\frac{Bu_s}{2\lambda}+q\lambda\Big).
\end{equation}
\noindent
The boundary condition $u(2\lambda) = u_s$ and equation (\ref{eqn:conserve-flux})
will provide a relationship between $u_s$ and $u_m$ through $\lambda$.

For $d=2$, equation (\ref{eqn:ode-1}) is very much simplified, and  we find 
\begin{equation}
  \label{eqn:sol-2d}
  u(s)=u_m+\frac{(Bu_m+2q\lambda^2)\lambda^B e^{\lambda^2}\Gamma(-B/2,s^2/4)}
  {2-B\lambda^Be^{\lambda^2}\Gamma(-B/2,\lambda^2)},
\end{equation}
\noindent
where we utilized $u(\infty) = u_m$ and equation (\ref{eqn:conserve-flux}). Here
 $\Gamma(a,z)$ is the incomplete
gamma function defined by the integral $\Gamma(a,z)=\int_z^\infty
t^{a-1}e^{-t}dt$ \cite{Abramowitz_Stegun_1964}. The yet unknown parameter
$\lambda$ is found from relation (\ref{eqn:sol-2d}) at $u(2\lambda) =
u_s$ for a set of input parameters $u_s$, $u_m$ $q$, and $B$, through which the concentration
field, equation (\ref{eqn:sol-2d}), and the growth of second-phase
($R=2\lambda\sqrt{Dt}$) are determined.

Let us consider the case of $d=3$, that is assume that the
potential in equation (\ref{eqn:screw}) is meaningful for a
spherically symmetric system. In this case, for $B \ne 0$, the point $z=0$ is
a regular singularity of  equation
(\ref{eqn:ode-1}), while $z=\infty$ is an irregular singularity for
this equation, see appendix \ref{sec:appa} for further
consideration. Nevertheless, for
$d=3$, the general solution of equation (\ref{eqn:ode-1}) is expressed in the form
\begin{equation}
  \label{eqn:sol-3d}
 u(s)=
2C_1\,{_1\!F}_1\Big(-\frac{1}{2};\frac{1+B}{2};-\frac{s^2}{4}\Big)s^{-1}
+ 2^{B}C_2\,{_1\!F}_1\Big(-\frac{B}{2};\frac{3-B}{2};-\frac{s^2}{4}\Big) s^{-B},
\end{equation}
\noindent
where  ${_1\!F}_1(a;b;z)$ is Kummer's confluent hypergeomtric function,
sometimes denoted by $M(a,b,z)$ \cite{Abramowitz_Stegun_1964}. The
integration constants $C_1$ and $C_2$ in equation (\ref{eqn:sol-3d}) can be determined
by invoking  equation (\ref{eqn:conserve-flux}) and also the condition
$u(\infty)=u_m$, cf. appendix \ref{sec:appa}.

\section{Computations}
\label{sec:comp}

To study the growth behavior of a second-phase in a solid solution
under the action of screw dislocation field, we attempt to compute the
growth rate constant as a function of the supersaturation parameter $k$,
defined as $k \equiv (u_s-u_m)/q_u$ with $q_u=u_p-u_s$, where $u_p$
 is the composition of the nucleus \cite{Aaron_et_al_1970}. For $d=2$,
i.e., a cylindrical second-phase platelet, equation (\ref{eqn:sol-2d}) with
$u(2\lambda) = u_s$ yields

\begin{equation}
  k  = \Bigg[\frac{2\lambda^2+Bu_m(u_p-u_s)^{-1}}{2-B\lambda^B e^{\lambda^2}\Gamma\big(-B/2,\lambda^2\big)}\Bigg]
  \lambda^B e^{\lambda^2}\Gamma\big(-B/2,\lambda^2\big).
\label{eqn:supsat-2d-exact}
\end{equation}
\noindent
For $B=0$, the relations obtained by Frank \cite{Frank_1950} are recovered,
namely
\begin{eqnarray}
 u(z)  & =  & u_m + q_u \lambda^2e^{\lambda^2}E_1(z^2/4), \label{eqn:sol-0d2} \\
 k & = & \lambda^2 e^{\lambda^2}E_1(\lambda^2), \label{eqn:flux-0d2}
\end{eqnarray}
\noindent
where $E_1(x)$ is the exponential integral of order one, related to the
incomplete gamma function through the identity
$E_n(x)=x^{n-1}\Gamma(1-n,x)$ \cite{Abramowitz_Stegun_1964}.

From equation (\ref{eqn:supsat-2d-exact}), it is seen that a complete separation of the supersaturation parameter
$k \equiv (u_s-u_m)(u_p-u_s)^{-1}$ is not possible for $B \ne
0$. However, for $u_s << u_p$ (a reasonable proviso) we write
\begin{equation}
  k  = \Big(\lambda^2 + \frac{B}{2}\,\epsilon\Big)\, 
  \lambda^B e^{\lambda^2}\Gamma\big(-B/2,\lambda^2\big)+\mathcal{O}(\epsilon^2),
\label{eqn:supsat-2d-approx}
\end{equation}
\noindent
with $\epsilon \equiv u_s/u_p$. For $B=1$, equations
(\ref{eqn:sol-2d}) and (\ref{eqn:supsat-2d-approx}) yield, respectively
\begin{eqnarray}
 u(z)  & =  & u_m + \frac{2\lambda\,e^{\lambda^2}\,(u_m + 2q_u\lambda^2)\,E_{3/2}(z^2/4)}
{[2-e^{\lambda^2} E_{3/2}(\lambda^2)]z},
 \label{eqn:sol-1d2} \\
 k & = & \Big(\lambda^2 + \frac{\epsilon}{2}\Big)\,e^{\lambda^2}E_{3/2}(\lambda^2)+\mathcal{O}(\epsilon^2).
\label{eqn:flux-1d2}
\end{eqnarray}
\noindent
Similarly for $B=2$, we have 
\begin{eqnarray}
 u(z)  & =  & u_m + \frac{4\lambda^2e^{\lambda^2}(u_m + q_u \lambda^2)E_2(z^2/4)
}{[1-e^{\lambda^2} E_{2}(\lambda^2)]z^2},
 \label{eqn:sol-2d2} \\
 k & = & (\lambda^2 + \epsilon)E_2(\lambda^2)+\mathcal{O}(\epsilon^2).
\label{eqn:flux-2d2}
\end{eqnarray}
\noindent

\begin{figure}[htbp]
\begin{center}
    \includegraphics[width=0.80\textwidth]{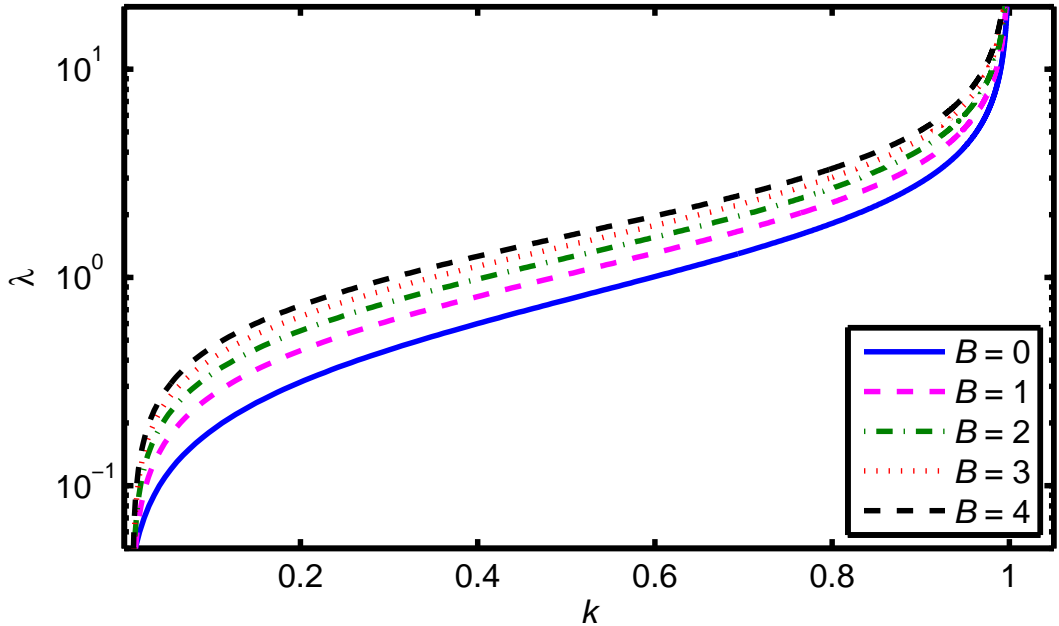}
 \caption{ Growth coefficient $\lambda$ as a function of supersaturation $k$ 
   at various levels of dislocation force
amplitude $B$ for a circular plate ($d=2$) and $u_s=0.01u_p$.}
\label{fig:k-2d}
  \end{center}
\end{figure}

\begin{figure}[htbp]
 \begin{center}
    \includegraphics[width=0.80\textwidth]{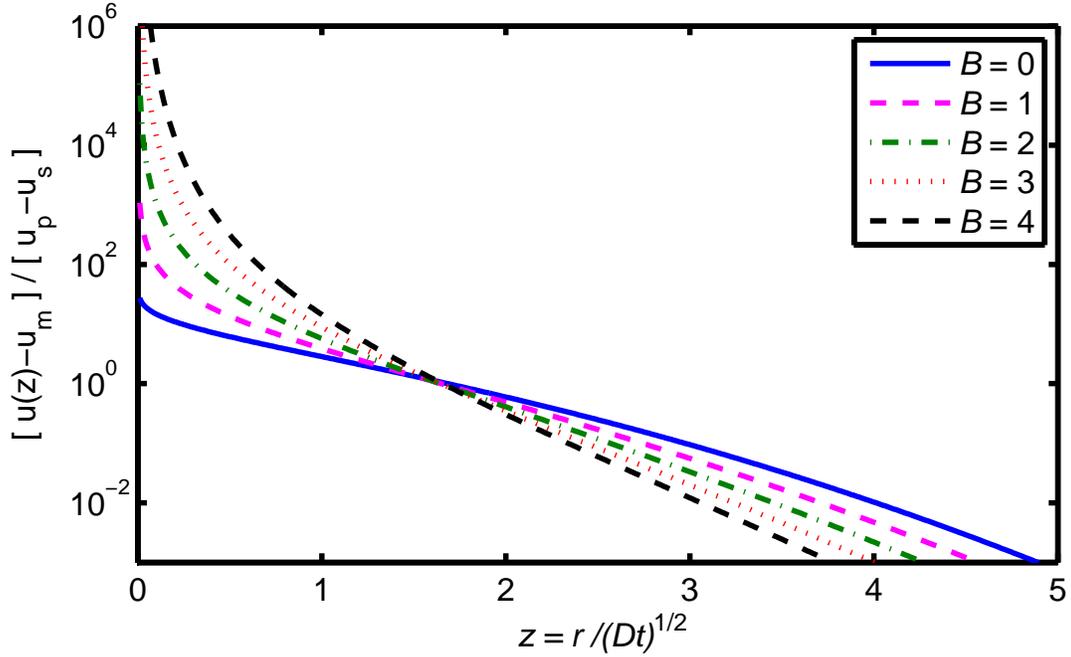}
\caption{Reduced concentration field as a function of reduced distance
 from the surface of the circular plate ($d=2$) at various levels of dislocation force
amplitude $B$ and at $\lambda=1$.}
\label{fig:u-2d}
\end{center}
\end{figure}

We have plotted the growth coefficient $\lambda=R/2\sqrt{Dt}$ as a
function of the supersaturation parameter $k$ in figure
\ref{fig:k-2d} and the spatial variation of the concentration field
in figure \ref{fig:u-2d} for $d=2$ and several values of $B$. The
computations are performed to $\mathcal{O}(\epsilon^2)$ with
$\epsilon=0.01$. Figure \ref{fig:k-2d} shows that $\lambda$ is an
increasing function of $k$; and also, as $B$ is raised $\lambda$ is
elevated. This means that an increase in the amplitude of dislocation
force (e.g., the magnitude of the Burgers vector) enhances second-phase
growth in an alloy.

Figure \ref{fig:u-2d} displays the reduced concentration versus the
reduced radius $z=r/\sqrt{Dt}$ for $\lambda=1$. The reduced
concentration is calculated via equation (\ref{eqn:sol-2d}). It is seen that for
$z \lesssim 1.6$ the concentration is enriched with increase in $B$,
whereas for $z \gtrsim 1.6$, it is vice versa. So, for
$\lambda=1$, the crossover $z$-value is $z_c \approx 1.6$. Also, as
$\lambda$ is reduced, $z_c$ is decreased.


For $d=3$, i.e., a spherical second-phase particle in the absence of
dislocation field ($B=0$), we find
\begin{eqnarray}
 u(z)  & =  & u_m + 2 q_u
\lambda^3 e^{\lambda^2}\Big[\frac{2e^{-z^2/4}}{z}
-\sqrt{\pi}\,\mathrm{erfc}(z/2)\Big],
 \label{eqn:sol-3d-b0} \\
 k & = & 2\lambda^2\Big[1-\sqrt{\pi}\,\lambda\, e^{\lambda^2}\mathrm{erfc}(\lambda)\Big].
\label{eqn:flux-3d-b0}
\end{eqnarray}
\noindent
This corresponds to the results obtained by Frank
\cite{Frank_1950}.

 For $d=3$ and $B=2$, equation (\ref{eqn:ode-1}) is
simplified and an analytical solution can be found, resulting in
\begin{eqnarray}
 u(z) & = & \Bigg(\frac{e^{z^2/4}(z^2+2)\Big[\sqrt{\pi}\lambda
e^{\lambda^2}\Big(\mathrm{erf}(\frac{z}{2})-\mathrm{erf}(\lambda)\Big)-1\Big]
+2\lambda e^{\lambda^2}z}{\sqrt{\pi}\lambda e^{\lambda^2}\mathrm{erfc}(\lambda)-1}\Bigg)
\frac{e^{-z^2/4}}{z^2}u_m +
\nonumber\\
 & & + \;
 \Bigg(\frac{\lambda^3 e^{\lambda^2}\Big[2z-\sqrt{\pi}e^{z^2/4}(z^2+2)\mathrm{erfc}(\frac{z}{2})\Big]}
{\sqrt{\pi}\lambda e^{\lambda^2}\mathrm{erfc}(\lambda)-1}\Bigg)\frac{e^{-z^2/4}}{z^2}q_u.
 \label{eqn:sol-3d2}
\end{eqnarray}
\noindent
Putting $u(2\lambda)=u_s$, we obtain
\begin{equation}
 k   =  \frac{1+2\lambda^2\Big(1-\sqrt{\pi}\lambda\,e^{\lambda^2}\mathrm{erfc}(\lambda)\Big)}
{2\lambda^2\Big(\sqrt{\pi}\lambda\,e^{\lambda^2}\mathrm{erfc}(\lambda)-1\Big)}\;\frac{u_m}{q_u}+
\frac{2\lambda^2-(1+2\lambda^2)\sqrt{\pi}\lambda\,e^{\lambda^2}\mathrm{erfc}(\lambda)}
{2\Big(\sqrt{\pi}\lambda\,e^{\lambda^2}\mathrm{erfc}(\lambda)-1\Big)}.
\label{eqn:k-3d-2e}
\end{equation}
For $u_s << u_p$,  we write
\begin{equation}
 k  =
-2\lambda^4+\sqrt{\pi}\lambda^3(1+2\lambda^2)\,e^{\lambda^2}\mathrm{erfc}(\lambda)
 + \Big(1-2\lambda^2 + 2\sqrt{\pi}\lambda^3\,e^{\lambda^2}\mathrm{erfc}(\lambda)\Big)\epsilon
+\mathcal{O}(\epsilon^2).
\label{eqn:k-3d-2a}
\end{equation}
\noindent

General analytical expressions of $u(z)$ and $k$, in terms of confluent
hypergeometric functions, can also be found for even values of $B$ as
detailed in appendix \ref{sec:appa}. Furthermore, asymptotic forms of
$u(z)$ for large and small $z$ can be calculated, see appendix
\ref{sec:appa} for analysis of $z>>1$. Figure \ref{fig:k-23d} compares $k$
versus $\lambda$ for $d=2$ and $d=3$ in the absence of dislocation
field ($B=0$).

\begin{figure}[htbp]
\begin{center}
    \includegraphics[width=0.80\textwidth]{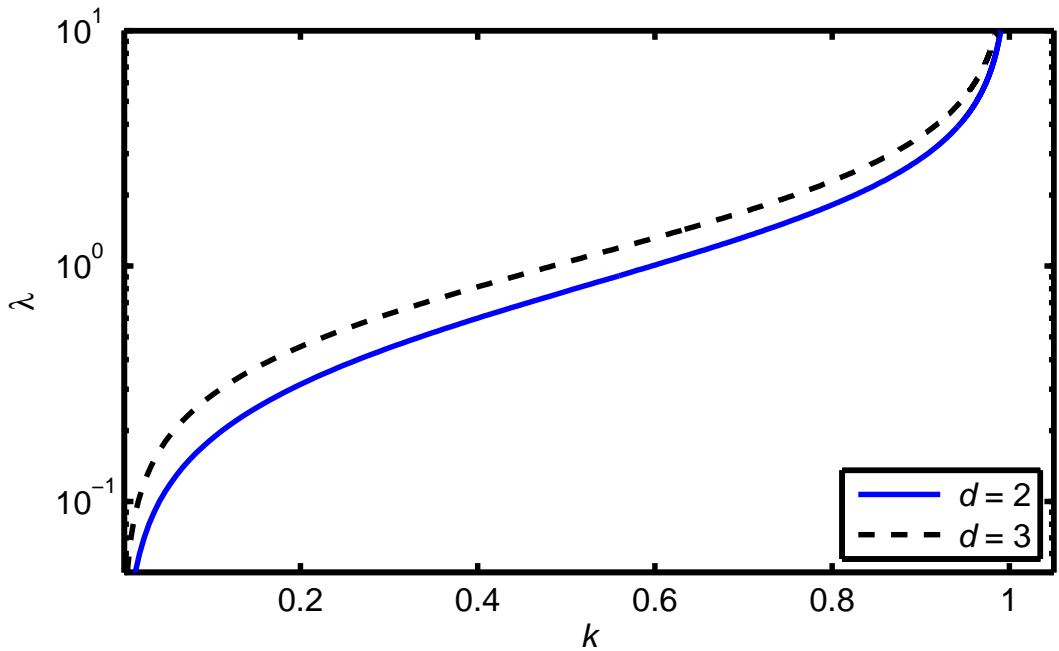}
 \caption{Growth coefficient $\lambda$ as a function of supersaturation parameter $k$ 
  at $B=0$ for a circular plate ($d=2$) versus a sphere ($d=3$).}
\label{fig:k-23d}
  \end{center}
\end{figure}

\section{Discussion}
 \label{sec:disc}

The potential energy in equation (\ref{eqn:screw}) describes the
elastic energy of the dislocation relaxed within the volume
occupied by the second-phase precipitate \cite{Cahn_1957}. It was treated
here as an external field affecting the diffusion-limited growth of
second-phase precipitate. The interaction energy of impurities in a
crystalline with dislocations depends on the specific model or
configuration of a solute atom and a matrix which is used. Commonly,
it is assumed that the solute acts as an elastic center of
dilatation. It is a fictitious sphere of radius $R^\prime$ embedded
concentrically in a spherical hole of radius $R$ cut in the matrix. If
the elastic constants of the solute and matrix are the same, the work
done in inserting the atom in the presence of dislocation is
$w=p\Delta v$, where $p$ is the hydrostatic pressure and $\Delta v$ is
the difference between the volume of the hole in the matrix and the
sphere of the fictitious impurity. For a screw dislocation $p=0$,
while near an edge dislocation
$p=\frac{(1+\nu)bG\sin\theta}{3\pi(1-\nu)r}$ for an impurity with
polar coordinates $(r,\theta)$ with respect to the dislocation $0z$,
hence $w \propto \Delta v\sin\theta/r$
\cite{Cottrell_Bilby_1949}. Using a nonlinear elastic theory
\cite{Nabarro_1987}, a screw dislocation may also interact with the
spherical impurity with the interaction energy $w \propto \Delta
v/r^2$. Moreover, accounting for the differences in the elastic
constants of a solute and a matrix, the solute will relieve shear
strain energy as well as dilatation energy, which will also interact
with a screw dislocation with a potential $w \propto \Delta v/r^2$
\cite{Friedel_1967}. Indeed, Friedel \cite{Friedel_1967} has
formulated that by introducing a dislocation into a solid solution of
uniform concentration $c_0$, the interaction energy between the
dislocation and solute atoms can be written as $w \backsimeq
w_0(b/\delta)^n f(\theta)$, where $\delta$ is the distance between the two
defects, $w_0$ the binding energy when $\delta=b$, and $f(\theta)$
accounts for the angular dependence of the interaction along the
dislocation. Also, $n=1$ for size effects and $n=2$ for effects due to
differences in elastic constants.  The discussed model for the
interaction energy between solute atoms and dislocations has been used
to study the precipitation process on dislocations by number of
workers in the past \cite{Ham_1959,Bullough_Newman_1962} and thoroughly
reviewed in \cite{Bullough_Newman_1970}. These studies concern
primarily the overall phase transformation (precipitation of a new
phase) rather than the growth of a new phase considered in our
note. That is, they used different boundary conditions as compared to the
ones used here.

Let us now link the supersaturation parameter $k$ to an experimental
situation. For this purpose, the values of $u_s$, i.e. the
concentration at the interface between the second-phase and matrix
should be known. The capillary effect leads to a relationship between
$u_s$ and the equilibrium composition $u_{eq}$ (solubility line in a
phase diagram). To obtain this relationship, we consider an incoherent
nucleation of second-phase on a dislocation \`a la Cahn
\cite{Cahn_1957}. A Burgers loop around the dislocation in the matrix
material around the incoherent second-phase (circular plate) will have
a closure mismatch equal to $b$. Following Cahn, on forming the
incoherent plate of radius $R$, the total free energy change per unit
length is
\begin{equation}
\label{eqn:cahn-fe}
\mathcal{G}=-\pi R^2 \Delta g_v +2\pi\gamma
R-A^\prime\ln(R/r_0),
\end{equation}
\noindent
where $\Delta g_v$ is the volume free energy of
formation, $\gamma$ the interfacial energy and the last term is the
dislocation energy, $A^\prime=Gb^2/4\pi$ for screw dislocations,
cf. equation (\ref{eqn:screw}).  Setting ${\rm d}\mathcal{G}/{\rm d}R=0$, yields
\begin{equation}
  \label{eqn:crit-rad}
  R = \frac{\gamma}{2\Delta g_v}\Big(1 \pm \sqrt{1-\alpha}\Big),
\end{equation}
\noindent
where $\alpha=2A^\prime \Delta g_v/\pi\gamma^2$. So, if $\alpha>1$,
the nucleation is barrierless, i.e., the phase transition kinetics is
only governed by growth kinetics, which is the subject of our
investigation here. If, however, $\alpha < 1$, there is an energy
barrier and the local minimum of $\mathcal{G}$ at $R=R_0$, which
corresponds to the negative sign in equation (\ref{eqn:crit-rad}),
ensued by a maximum at $R=R^\ast$ corresponding to the positive sign
in this equation. The local minimum corresponds to a subcritcal
metastable particle of the second-phase surrounding the dislocation
line, and it is similar to the Cottrell atmosphere of solute atoms in
a segregation problem. When $\alpha = 0$, corresponding to $B=0$, the
two phases are in equilibrium and the maximum in $\mathcal{G}$ is
infinite, as for homogeneous nucleation.

\begin{figure}[htbp]    
\begin{center}
\includegraphics[width=0.80\textwidth]{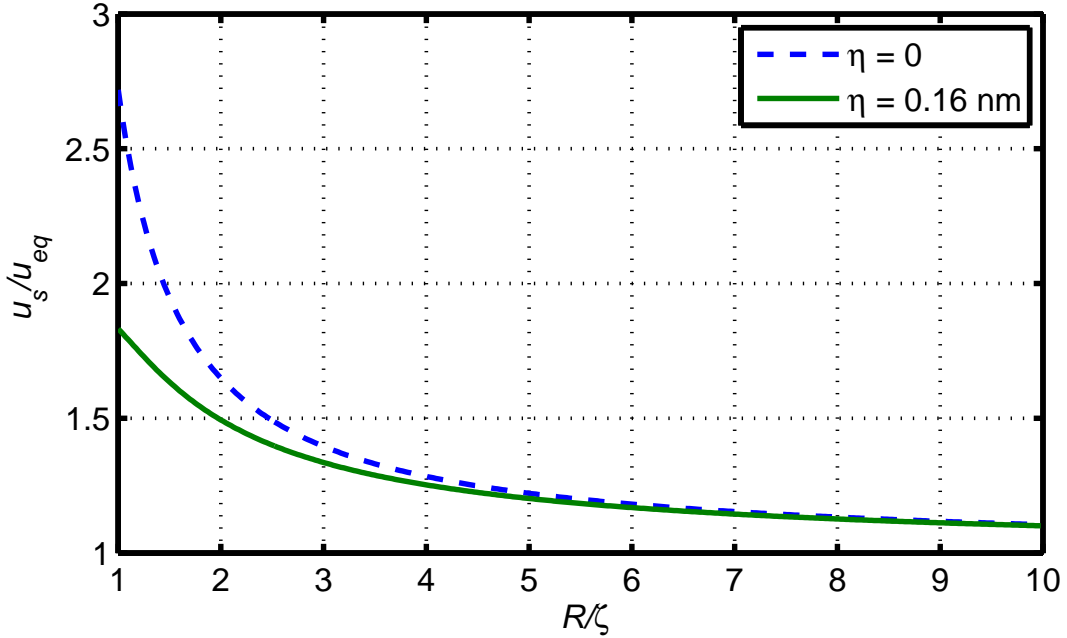}{} \\
\caption{The size dependence of the concentration at the curved
precipitate/matrix interface $u_s$ relative to that of the flat
interface $u_{eq}$ for a set of parameter values given in the
text, cf. eq. (\ref{eqn:gibbs-thom}).}
\label{fig:thoms-2d}
\end{center}
\end{figure}

For a dilute regular solution, $\Delta
g_v=(k_BT/V_p)\ln(u_s/u_{eq})$, where $V_p$ is the atomic volume of
the precipitate compound, $u_s$ is the concentration of the matrix at
a curved particle/matrix interface and $u_{eq}$ that of a flat
interface, which is in equilibrium with the solute concentration in
the matrix. Equation (\ref{eqn:crit-rad}) gives
$\Delta g_v=\gamma/R-A^\prime/2\pi R^2$. Hence, for a dilute regular
solution, we write
\begin{equation}
  \label{eqn:gibbs-thom}
 u_s=u_{eq}\exp{\Big[\frac{\zeta}{R}\Big(1-\frac{\eta}{R}\Big)\Big]},
\end{equation}
\noindent
where $\zeta = \beta V_p\gamma$, $\beta=1/k_BT$ and $\eta = A^\prime/2\pi\gamma$. Subsequently, the
supersaturation parameter is expressed by
\begin{equation}
  \label{eqn:supsat}
  k = \frac{u_{eq}\exp[{\frac{\zeta}{R}(1-\frac{\eta}{R})}]-u_m}{u_p-u_{eq}\exp[\frac{\zeta}{R}(1-\frac{\eta}{R})]}.
\end{equation}
\noindent
Taking the following typical values: $\gamma=0.2$ Jm$^{-2}$, $G=40$
GPa, and $b=0.25$ nm, then $A^\prime \approx 2.0\times10^{-10}$ N and
$\eta=0.16$ nm. Figure \ref{fig:thoms-2d} depicts $u_s/u_{eq}$, from
equation (\ref{eqn:gibbs-thom}), as a function of scaled radius
$R/\zeta$ for $V_p=1.66 \times 10^{-29}$ m$^3$, $\eta=0$ and
$\eta=0.16$ nm at $T=600$ K. Equation
(\ref{eqn:gibbs-thom}) is analogous to the Gibbs-Thomson-Freundlich
relationship \cite{Christian_2002} comprising a dislocation defect.

Recalling now the values used for the interaction parameter $B$ in the
computations presented in the foregoing section, we note that for
$B=2$ and the above numerical values for $G$ and $b$ at  $T=1000$ K, we
find $l\approx 0.14$ nm, which is close to the calculated value of
$\eta$.

In Cahn's model, the assumption that all the strain energy of the
dislocation within the volume occupied by the nucleus can be relaxed
to zero demands that the nucleus is incoherent. For a coherent nucleus
forming on or in proximity of dislocations, this supposition is not
true. Instead, it is necessary to calculate the elastic interaction
energy between the nucleus and the matrix, which for an edge
dislocation is in the form $Gb^2/[4\pi(1-\nu)r]$ for the energy
density per unit length \cite{Barnett_1971}. In the same manner, to
extend our calculations for growth of coherent precipitate, we must
employ this kind of potential energy, i.e. the potential energy of the
form $\phi(r)=-A\ln(r/r_0)+C \sin\theta/r$, in the governing kinetic
equation rather than relation (\ref{eqn:screw}).

\appendix
\section{Evaluation of solutions of equation (\ref{eqn:ode-1}) for $d=3$}
\label{sec:appa}
For an ordinary second-order differential equation with a regular
singularity, the Frobenius method can be used to obtain power series
solution. On the other hand, when singularity is irregular, no
convergent solution may be found; nevertheless, albeit divergent, the
solution can be asymptotic. Let us write equation (\ref{eqn:ode-1})
for $d=3$ in a generic form
\begin{equation}
  \label{eqn:sod-gen1}
  u''+ p(z)u'+ q(z)u = 0,
\end{equation}
\noindent
where primes denote differentiation with respect to $z$,
$p(z)=z/2+(2+B)/z$ and $q(z)=B/z^2$. Since we have imposed the
boundary condition $u(\infty)=u_m$, it is worthwhile to explore the
behavior of the solution as $z\to\infty$. But, first let us put equation
(\ref{eqn:sod-gen1}) in a more convenient form by setting
$u(z)=\widetilde u (z)\exp[-\frac{1}{2}\int p(z)dz]$, which gives
$\widetilde u''+ (q-p'/2+p^4)\widetilde u = 0$. Here, without loss of
generality, we consider
\begin{eqnarray}
 \label{eqn:sod-gen2}
 u''+ f(z)u  & = & 0,\\\label{eqn:fz}
\text{where} \qquad f(z) & = & -\frac{z^2}{16} -\frac{B+3}{4}-\frac{B(B-2)}{4z^2}.
\end{eqnarray}
\noindent
 Since $f(z)$ is not $\mathcal{O}(z^{-2})$ as $z\to\infty$, then the
point at infinity is an irregular singularity for $u(z)$. We now look
for solutions of (\ref{eqn:sod-gen2}) by considering
\begin{equation}
 \label{eqn:sod-gen-sol}
 u(z)\sim \exp\Big(\sum_{n=0}^\infty\psi_n(z)\Big),
\end{equation}
\noindent
where $\{\psi_n(z)\},n=0,1,\dots$, is an asymptotic sequence as
$z\to\infty$. Substituting (\ref{eqn:sod-gen-sol}) into equation
(\ref{eqn:sod-gen2})
\begin{equation}
 \label{eqn:sod-3}
 \psi_0''+ \psi_1'' + \dots +(\psi_0'+ \psi_1'+\dots)^2 +f(z) \sim 0,
\end{equation}
\noindent
where we have tacitly assumed that $\psi_n(z)$ is (twice)
differentiable and the resulting series are still asymptotic.

Equation (\ref{eqn:sod-3}) is used to determine the
$\{\psi_n(z)\},n=0,1,\dots$ by successively applying the asymptotic
limit $z\to\infty$. The results for the first few terms are shown in
table \ref{tab:sols}. Hence, we write for $z\to\infty$:
\begin{eqnarray}
 \label{eqn:sod3-sol1}
u_+(z) & \sim & A_1e^{z^2/8}z^{(1+B/2)}\Big[1+\frac{B}{z^2} + \mathcal{O}(z^{-4})\Big],\\
\label{eqn:sod3-sol2}
u_-(z) & \sim & A_2e^{-z^2/8}z^{-(B/2+2)}\Big[1-\frac{3(B+2)}{z^2} + \mathcal{O}(z^{-4})\Big],
\end{eqnarray}
\noindent
where $A_1$ and $A_2$ are arbitrary constants. Note that the solution
(\ref{eqn:sod3-sol1}) is divergent for large $z$, whereas
(\ref{eqn:sod3-sol2}) is convergent and thus is physically
admissible. Considering $u(\infty)=u_m$, we write
\begin{equation}
 \label{eqn:sod-4}
u_-(z)  \sim  u_m +
A_2e^{-z^2/8}z^{-B/2-2}\Big[1-\frac{3(B+2)}{z^2} +
\mathcal{O}(z^{-4})\Big],\; \text{as}\; z \to \infty.
\end{equation}
\noindent

\begin{table}
  \caption{\label{tab:sols}Solutions to equation (\ref{eqn:sod-3}).}
{\begin{tabular}{@{}lcc}\toprule
   Sequence   & Solution 1 & Solution 2  \\
\colrule
   $\psi_0$ & $z^2/8$ & $-z^2/8$   \\
    $\psi_1$ & 0 & 0   \\
    $\psi_2$ & $(\frac{B+2}{2})\ln z$  &  $-(\frac{B+4}{2})\ln z$   \\
    $\psi_3$ & 0 & 0   \\
    $\psi_4$ & $Bz^{-2}$  &  $-3(B+2)z^{-2}$   \\
   \botrule
  \end{tabular}}
\end{table}

Let us now evaluate the general solution to equation (\ref{eqn:ode-1})
for $d=3$ as expressed by equation (\ref{eqn:sol-3d}). We apply the
flux conservation relation (\ref{eqn:flux-1}) to obtain $C_1$, and
then substitute $C_1$ in equation (\ref{eqn:sol-3d}) to write

\begin{equation}
\label{eqn:sol-3d-gen}
u(z)=K_1(z,B)C_2+K_2(z,B)q,
\end{equation}
where 
\begin{eqnarray}
\label{eqn:sol3d-1}
K_1(z,B) & = & \Big(\frac{2}{z}\Big)^B 
{_1\! F}_1\Big(-\frac{B}{2};\frac{3-B}{2};-\frac{z^2}{4}\Big)+\nonumber \\
& & +
\frac{B\lambda^{-B}{_1\! F}_1\Big(\frac{2-B}{2};\frac{5-B}{2};-\lambda^2\Big)}{B-3}K_2(z,B),\\
\label{eqn:sol3d-2}
K_2(z,B) & = &
\frac{4(B+1)\lambda^3{_1\!F}_1(-\frac{1}{2};\frac{1+B}{2};-\frac{z^2}{4})/z}
{(B^2-1){_1\!F}_1(-\frac{1}{2};\frac{1+B}{2};-\lambda^2)+2\lambda^2{_1\!F}_1(\frac{1}{2};\frac{3+B}{2};-\lambda^2)},\\
\label{eqn:sol3d-3}
 K_2(z,0)  & =  &   2 \lambda^3 e^{\lambda^2}\Big[\frac{2e^{-z^2/4}}{z}
+\sqrt{\pi}\,\mathrm{erf}(z/2)\Big].
\end{eqnarray}
\noindent
Here, ${_1\!F}_1(a;b;z)$ is the confluent hypergeometric function. If
$a<0$, and either $b>0$ or $b<a$, this function can be expressed as a
polynomial with finite number of terms. If, however, $b=0$ or a
negative integer, then ${_1\!F}_1(a;b;z)$ itself is infinite. Thus,
relations (\ref{eqn:sol3d-2})-(\ref{eqn:sol3d-3}) become singular for
$B=1,3,5,7,\dots$, making the solutions meaningless. Some useful
relations for computations are listed in table
\ref{tab:identities}. Additional relations and properties for
${_1\!F}_1(a;b;z)$ can be found in \cite{Abramowitz_Stegun_1964}.

Next, we utilize the remote boundary condition $u(\infty)=u_m$ to
determine $C_2$, then we formally write
\begin{equation}
\label{eqn:sol-3d-gen2}
u(z)=\frac{K_1(z,B)}{K_1(\infty,B)}u_m + \Big(K_2(z,B)-\frac{K_1(z,B)}{K_1(\infty,B)}K_2(\infty,B)\Big)q.
\end{equation}
\noindent
 In computations of $K_1(\infty,B)$ prudence must be exercised, i.e.,
first evaluate this quantity for a given value of $B$, then take the
limit $z \to \infty$. Note also that $K_1(z,0)=1$ $\forall z$ and
$K_2(\infty,0)=2\sqrt{\pi}\lambda^3 e^{\lambda^2}$.

Furthermore, we may calculate a relation for the supersaturation
parameter, $k=(u_s-u_m)/(u_p-u_s)$, defined in the main text by using the condition
$u(2\lambda)=u_s$ on equation (\ref{eqn:sol-3d-gen}), which gives
\begin{equation}
\label{eqn:k-3d-gen}
 k = \Bigg(\frac{K_1(2\lambda,B)}{K_1(\infty,B)}-1\Bigg)\frac{u_m}{q} +
  K_2(2\lambda,B)-\frac{K_1(2\lambda,B)}{K_1(\infty,B)}K_2(\infty,B).
\end{equation}
For dilute alloys, $u_s<<u_p$; so with $\epsilon \equiv u_s/u_p$, we write
\begin{equation}
\label{eqn:k-3d-dil}
k=\Bigg(1-\frac{K_1(\infty,B)}{K_1(2\lambda,B)}\Bigg)\epsilon +
  \frac{K_2(2\lambda,B)}{K_1(2\lambda,B)}K_1(\infty,B) - K_2(\infty,B)+ \mathcal{O}(\epsilon^2).
\end{equation}
\begin{table}
  \caption{\label{tab:identities}Special cases of  ${_1\!F}_1(a;b;z)$.}
{\begin{tabular}{@{}l}\toprule
   ${_1\!F}_1(-1;\frac{1}{2};-z^2) = 1+2z^2$ \\
  ${_1\!F}_1(-\frac{1}{2};\frac{1}{2};-z^2)=e^{-z^2}+\sqrt{\pi}z\,\mathrm{erf}(z)$\\
  ${_1\!F}_1(-\frac{1}{2};\frac{3}{2};-z^2)=\frac{1}{2}e^{-z^2}+\frac{\sqrt{\pi}}{4}(1+2z^2)\mathrm{erf}(z)z^{-1}$\\
  ${_1\!F}_1(\frac{1}{2};\frac{3}{2};-z^2)=\frac{\sqrt{\pi}}{2}\mathrm{erf}(z)z^{-1}$\\
  ${_1\!F}_1(\frac{1}{2};\frac{5}{2};-z^2)=\frac{3}{8}\Big(2z e^{-z^2}+\sqrt{\pi}(2z^2-1)\mathrm{erf}(z)\Big)z^{-3}$
\\\hline\hline
  \end{tabular}}
\end{table}

\bibliographystyle{apsrev} \bibliography{micro}

\end{document}